\documentclass[reprint,twocolumn, aps, prl]{revtex4-1}

\usepackage{graphicx,psfrag}
\usepackage{epstopdf}
\usepackage{amsmath}
\usepackage{bm}
\usepackage{amssymb}
\usepackage{subfigure}

\newcommand{\pdone}[2]{{\frac{\partial #1}{\partial #2}}}

\begin{document}

\title{\large \bf Phononic Thin Plates with Embedded Acoustic Black Holes}

\author{Hongfei Zhu}
\email{Hongfei.Zhu.44@nd.edu}
\author {Fabio Semperlotti}
\email{Fabio.Semperlotti.1@nd.edu}
\affiliation{Department of Aerospace and Mechanical Engineering, University of Notre Dame, Notre Dame, IN 46556}

\begin{abstract}
We introduce a class of two-dimensional non-resonant single-phase phononic materials and investigate its peculiar dispersion characteristics. The material consists of a thin plate-like structure with an embedded periodic lattice of Acoustic Black Holes. The use of these periodic tapers allows achieving remarkable dispersion properties such as Zero Group Velocity in the fundamental modes, negative group refraction index, bi-refraction, and mode anisotropy. The dispersion properties are numerically investigated using a three-dimensional supercell plane wave expansion method. The effect on the dispersion characteristics of key geometric parameters of the black hole, such as the taper profile and the residual thickness, are also explored.
\end{abstract}

\maketitle

 Phononic Crystals (PC) are artificial media made of two or more materials combined together to form a periodic structure. These materials offer unusual wave propagation characteristics such as acoustic bandgaps \cite{Liu,Yang,Helios,Garcia,Vasseur}, localized and guided defect modes \cite{Torres,Kafesaki,Khelif}, filtering of acoustic waves \cite{Pennec1,Pennec2,Zhu}, acoustic lenses \cite{Zhu2,Semperlotti}, and negative refraction \cite{Morvan,Pierre}, that are typically not achievable in conventional materials. PCs are often classified in two categories, non-resonant and locally-resonant \cite{Cui2009}, in order to highlight the difference between their operating modes. The locally-resonant materials exhibit low frequency resonances (typically in the metamaterial range) localized at the inclusion while the non-resonant materials exhibit inclusion resonances only in the high frequency range. Owing to this mechanism, the wave propagation characteristics in the low frequency range are mostly (acoustic) impedance-driven for the non-resonant materials and inertia-driven for the locally resonant materials. These local resonances have been shown to be strictly related to the generation of negative effective properties (such as density and bulk modulus) which are at the basis of double negative properties \cite{Lee}.

\begin{figure}[h!]
  \center{\includegraphics[scale=0.4]{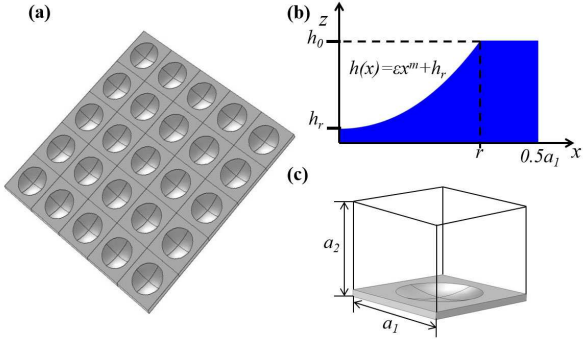}}
  \caption{(a) Schematic of the phononic thin plate with a squared ABH periodic lattice structure, (b) cross section of the acoustic black hole showing the taper profile, and (c) 3D supercell used for the PWE model.}\label{Fig1}
\end{figure}

 Despite such remarkable dynamic properties, the integration of these materials into practical devices and applications is still lacking. The fabrication complexity (particularly for the locally resonant type) and the non-structural character (i.e. not load bearing materials) of typical PC designs are among the main limiting factors. In this study, we propose a class of two-dimensional PCs obtained by tailoring the geometry of a single-phase isotropic material able to provide the same high-level characteristics of locally resonant PCs. These materials are synthesized by embedding a periodic lattice of carefully engineered geometric inhomogeneities consisting of tapered holes. These inhomogeneities can be introduced (virtually) in any material by simply manufacturing tapers having prescribed profiles in the host structure. Among the fabrication advantages of this design, we highlight that it does not require interfacing multiple materials and it can be retrofitted even to existing structures. This could have critical implications to develop highly absorbing thin-walled structures with embedded passive vibration and acoustic control capabilities. The proposed phononic structure (Fig.\ref{Fig1}) consists in a thin plate made of a periodic lattice of exponential-like circular tapers, often referred to as \textit{Acoustic Black Holes} (ABH).

 The physical principle exploited in ABHs was first observed by Pekeris \cite{Pekeris} for waves propagating in stratified fluids and later extended to acoustics in solids by Mironov \cite{Mironov}. Mironov observed that, under certain conditions, flexural waves propagating in a thin plate with an exponentially tapered edge will theoretically never reflect back, therefore resulting in the so-called \textit{zero reflection} condition. More recently Krylov \cite{Krylov1,Krylov2} exploited this concept to achieve passive vibration control of structural elements. The ABH consists in a variable thickness exponential-like circular taper able to produce a progressive reduction of the phase and group velocity as the wave approach the center of the hole. Typical thickness profiles are of the form $h(x)=\varepsilon x^m$ where $\{ m$, $\varepsilon \}$ $\in \mathbb{R}$, $m \geq 2$, and $\varepsilon \ll (3 \rho \omega^2 / E)^{1/2}$ to satisfy the smoothness criterion \cite{Mironov,Feurtado}. In ideal ABH tapers, where the thickness decreases to zero, the phase and group velocities tend to zero as they approach the center of the ABH. Under this condition, the wave never reaches the center of the hole therefore the reflection coefficient approaches zero (the wave is not reflected back) and the hole appears as an ideal absorber. Energy balance considerations show that, in the absence of damping, the center of the hole becomes a point of singularity for both the particle displacement \cite{Mironov} and the vibrational energy \cite{Zhao}.
In practice, the residual thickness at the center of the ABH cannot be made zero due to both fabrication and structural constraints. In the absence of damping the residual thickness can produce appreciable levels of reflected energy (up to 70 \% \cite{Mironov2}) even for a small residual thickness. Nevertheless, wave speed reduction will still take place.

In this study, we consider an infinite thin plate with a periodic distribution of  ABH-like tapers as shown in Figure \ref{Fig1}a. The plate has thickness $h=8 $ mm and a taper profile $h(x)=\varepsilon x^m + h_r$ where $h_r$ is the residual thickness. The lattice structure is assembled from a square unit cell as shown in Figure \ref{Fig1}b.

The elastodynamic response of the phononic thin plate is governed by the Navier's equations:
\begin{align}
\rho \pdone{^2 u_i}{t^2}=(C_{ijkl}u_{k,l})_{,j} \qquad i=1,2,3 \label{Navier}
\end{align}
where $\rho$ is the density, $C_{ijkl}$ is the stiffness tensor, and $u_i$ are the components of the displacement field. Eqn. (\ref{Navier}) is also subjected to traction free boundary conditions $T_{xz}=T_{yz}=T_{zz}=0$ at the upper and lower surfaces. This condition is applied at $z= h(x)$ on the bottom surface and at $z= h/2$ if $0 \leq x \leq r$ or at $z=- h/2$ if $r \leq x \leq 0.5 a_1$ on the top surface.

The dispersion relations are obtained solving eqns. (\ref{Navier}) using a three dimensional Supercell Plane Wave Expansion (PWE) approach \cite{3DPWE}. We define a unit cell as shown in Figure \ref{Fig1}c. The material is rendered periodic also in the out-of-plane direction $z$ by alternating the thin plate with vacuum layers. Under these conditions, the material properties can be approximated using a three dimensional Fourier series expansion. Bloch periodic boundaries are enforced along the in-plane directions to simulate an infinite plate.

In order to solve eqns. (\ref{Navier}) the position dependent density $\rho (\vec{r})$ and elastic coefficients $C_{ijkl} (\vec{r})$ are expanded in Fourier series using the reciprocity vector $\vec{G}=(G_x,G_y,G_z )$:

\begin{align}
C_{ijkl}(\vec{r})=\sum_G e^{i\vec{G}\bullet\vec{r}}{C_{ijkl}}_G \label{C}
\end{align}
and
\begin{align}
\rho(\vec{r})=\sum_G e^{i\vec{G}\bullet\vec{r}}{\rho}_G \label{rho}
\end{align}
where $\rho_G$ and ${C_{ijkl}}_G$ are the corresponding Fourier coefficients and are defined as:

\begin{align}
{C_{ijkl}}_G=\frac{1}{V}\int_V C_{ijkl}(\vec{r})e^{-i\vec{G}\bullet\vec{r}}dr^{3}
\end{align}
and
 \begin{align}
\rho_G=\frac{1}{V}\int_V \rho(\vec{r})e^{-i\vec{G}\bullet\vec{r}}dr^{3}
\end{align}

After using the Bloch theorem and expanding the displacement vector $\vec{u(x, y, z, t)}$ in Fourier series, we obtain:
\begin{align}
\vec{u}(\vec{r})=\sum_{G\prime} A_{G\prime}e^{i[(\vec{k}+\vec{G'})\bullet\vec{r}-\omega t]} \label{disp}
\end{align}
where $\vec{k}=(k_x,k_y,0)$ is the Bloch plane wave vector, $\omega$ is the circular frequency, and $A_{G'}$ is the amplitude of the displacement vector. Substituting eqns. (\ref{C}),(\ref{rho}), and (\ref{disp}) into eqn. (\ref{Navier}) and collecting terms, we obtain the $3n \times 3n$ set of equations:

\begin{align}
\begin{split}
\begin{pmatrix}
 C^{11}_{G,G'}  & C^{12}_{G,G'} & C^{13}_{G,G'}   \\
 C^{21}_{G,G'}  & C^{22}_{G,G'} & C^{23}_{G,G'}     \\
  C^{31}_{G,G'}  & C^{32}_{G,G'} & C^{33}_{G,G'}     \\
\end{pmatrix}
\begin{pmatrix}
A^{1}_{G'} \\
A^{2}_{G'} \\
A^{3}_{G'} \\
\end{pmatrix}
= \\
\omega^2
\begin{pmatrix}
 \rho_{G,G'}  & 0 & 0   \\
 0  & \rho_{G,G'} & 0     \\
  0  & 0 & \rho_{G,G'}     \\
\end{pmatrix}
& \begin{pmatrix}
A^{1}_{G'} \\
A^{2}_{G'} \\
A^{3}_{G'} \\
\end{pmatrix}
\end{split} \label{eigen}
\end{align}

where the $n \times n$ sub-matrices $C_{G,G'}$ are functions of the Bloch wave vector $\vec{k}$, the reciprocal lattice vectors $\vec{G}$, the circular frequency $\omega$, and the Fourier coefficients $\rho_G$ and ${C_{ijkl}}_G $. The detailed expression can be found in \cite{3DPWE}. Equation (\ref{eigen}) can be written in the form of an eigenvalue problem whose solution provides the eigenfrequencies and the eigenmodes of the system.

\begin{figure}[h!]
  \center{\includegraphics[scale=0.35]{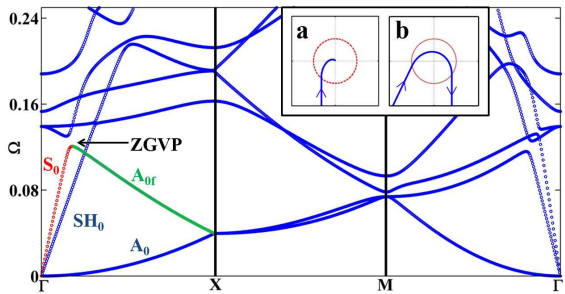}}
  \caption{Dispersion relation along the irreducible part of the first Brillouin zone for the phononic plate structure in reference configuration. The insets show the results of geometric acoustic analysis illustrating the effect of different tapers on an incoming ray. (a) corresponds to an ideal ABH (i.e.  $t \xrightarrow{} 0$) while (b) corresponds to the reference configuration.} \label{Fig2}
\end{figure}

In the following numerical study, we consider a reference configuration consisting in a 8 mm thick aluminum plate with tapers characterized by $m$=2.2, $\varepsilon$=5, radius $r$=0.05 m, and residual thickness $h_r$=0.0011m. The lattice has a squared configuration with lattice constant $a_1$=0.14 m.  The reciprocal lattice constants retained for the expansion are $G_x = G_y= \pm(3,2,1,0)2 \pi/a_1$ and $G_z =\pm(4,3,2,1,0)2\pi/a_2$, where $a_2$=0.064m. The constant $a_2$ was selected so to dynamically isolate the different slabs in the $z$ direction. The band structure along the boundary of the first Brillouin Zone (BZ) for normalized frequencies $\Omega= \frac{\omega a}{2\pi C_t} $ up to 0.25 is shown in Figure \ref{Fig2}.

The dispersion relations show several peculiar properties that are typically observable only in locally resonant materials. Several non-monotonous branches can be found in the low frequency range. In particular, for the fundamental non-monotonous modes, the constitutive branches are associated with different mode types and group velocity regions. As an example, the $S_0$ mode evolves into the $A_0$ after crossing a zero group velocity point (ZGVP) along the $\Gamma-X $ boundary. Similar behavior is observed for the $SH_0$ mode that evolves into a higher order flexural mode along the $\Gamma-X$ boundary. The ZGVP point also separates regions with positive and negative group velocity. It has long been known \cite{ZGVP} that the higher order Lamb modes in plates can display zero group velocity points corresponding to waves having finite phase velocity but vanishing group velocity. However, this behavior is quite unexpected for the fundamental modes. The ZGVP is related to the existence of a standing wave associated with a local resonance of the plate. The branch of the dispersion curve beyond the ZGVP is characterized by negative group velocity which corresponds to backward wave propagation. This phenomenon was never noticed on fundamental Lamb modes. The occurrence of the ZGVP and of the negative group velocity branch is related to the ability of the ABH cell to bend the wave in the direction of decreasing phase velocity gradient, that is towards the ABH center. Depending on the properties of the incoming wave and on the geometric characteristics of the taper, particularly on the taper exponent and the residual thickness, the wave can be either slowed down and captured by the ABH (Figure \ref{Fig2}, inset a) or bent in the backward direction (Figure \ref{Fig2}, inset b). This behavior was verified by performing a geometric acoustic analysis (insets in Figure \ref{Fig2}) to identify the trajectory of a ray traveling through the ABH. These two conditions can be related to the generation of a ZGVP and to backward propagation, respectively.

\begin{figure}[h!]
  \center{\includegraphics[scale=0.35]{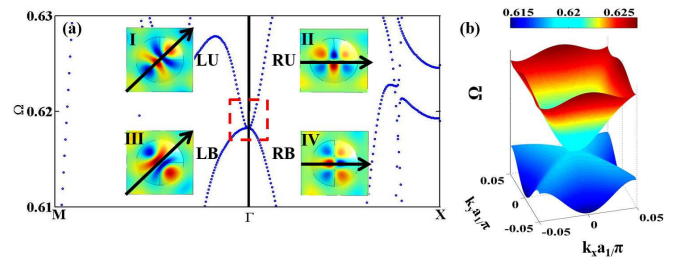}}
  \caption{(a) Zoom-in view of the dispersion relations around $\Omega =0.62$ showing the existence of a Dirac-point-like singularity. The insets I-IV show the modal displacements corresponding to the labeled modes LU,RU,LB and RB. The black arrow shows the wave vector direction. (b) shows the EFS plot around the DP-like singularity point.} \label{Fig3}
\end{figure}

Another interesting property of these metamaterials is the existence of several singularity points particularly at the $\Gamma$ and $X$ locations of the BZ. Each singularity point results from the intersection of upper and lower branches where only a degenerate mode at $\vec{k}=0$ exists. These points exhibit similar behavior to the well-known \textit{Dirac Points} (DP) \cite{Dirac}. An example of this Dirac-point-like singularity at $\Gamma$ is shown in Figure \ref{Fig3}. By analyzing the branches radiating outward from the singularity point(either along the $\Gamma-X$ or the $\Gamma-M$ boundary), we observe that they correspond to pairs of dissimilar flexural modes having different modal displacements (see insets in Fig \ref{Fig3}a). The modal displacement shape evolves from a $2\times3$ poles to a $4\times1$ poles as we transition from $\Gamma-X$ to $\Gamma-M$. More interestingly, if we follow the EFC for either the upper or lower branches, not only the modal displacement shape changes but also the axis of symmetry of the mode rotates. A close-up view of the Equi-Frequency-Surfaces around the DP-like singularity point is shown in Figure \ref{Fig3}b. We note that the shape of the lower branch clearly indicates anisotropic behavior. On the contrary, the upper branch (close to the DP point) is a quasi-circular cone, therefore suggesting quasi-isotropic characteristics in the selected frequency range. It is important to note that the cone is composed of different modes in different directions therefore indicating that the ABH-PC presents a \textit{"mode anisotropy"}. This concept is, in principle, equivalent to the super-anisotropy observed already in certain type of metamaterials \cite{Dirac}.

\begin{figure}[h!]
  \center{\includegraphics[scale=0.4]{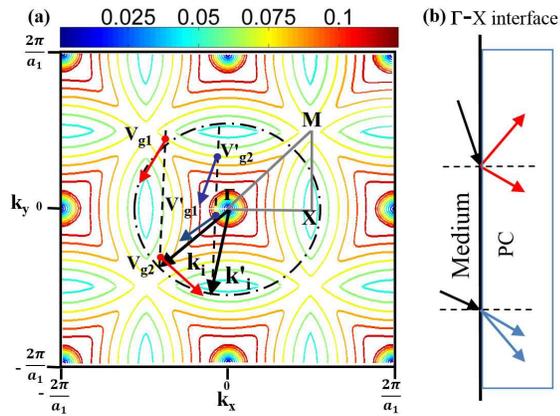}}
  \caption{(a) EFC for the fundamental non-monotonous mode $S_0-A_{0f}$. The black arrows and the black dashed lines indicate the incident angles and the wave vector conservation lines, respectively. The remaining arrows indicate the direction of the refracted beams. (b) Schematic of the bi-refraction mechanism.} \label{Fig4}
\end{figure}

To better understand the wave propagation produced by the ABH lattice structure, we extract the Equi-Frequency-Contours (EFC) for the fundamental non-monotonous mode $S_0-A_{0f}$ (Figure \ref{Fig4}). Results highlight the existence, in the same band, of a dual EFC contour associated with different group velocity directions. This aspect is of particular interest because it was shown in previous studies \cite{Bi-refra, Pichard} to be a fundamental condition for the existence of bi-refraction.

To illustrate this phenomenon we superimpose the EFC corresponding to an incident wave at a fixed frequency in the homogeneous (constant thickness) area of the plate (solid black circle). The wave vector of the refracted beam must satisfy the $k_{\parallel}$-conservation relation \cite{Pichard}: $k^{inc}_{\parallel}=k^{ref}_{\parallel}+G_{\parallel}$, where $k^{inc}_{\parallel}$ and $k^{ref}_{\parallel}$ are the components of the wave vector of the incident and refracted beam parallel to the interface, $G_{\parallel}$ is the parallel component of the reciprocal lattice vector. The group velocity is given by $V_g=\nabla\Omega(\vec{k})$ which is always perpendicular to the EFCs and pointing towards the direction of increasing frequency. In our case, dual EFCs with positive and negative group velocities co-exist at the same frequency, therefore bi-refraction should be expected. In particular, depending on the angle of incidence both positive-positive and positive-negative bi-refraction (see Figure \ref{Fig4}b) can be achieved. As shown in Figure \ref{Fig4}(a), $\Gamma-X$ is assumed as the interface boundary and the two black solid arrows represent two possible incident beams with different angle of incidence. The refracted beam are determined by finding the intersection point between the corresponding EFC and the conservation line (marked by the black dashed lines perpendicular to the $\Gamma-X$ boundary). Since the refracted beams are in the direction of the group velocity at the crossing point (red and blue arrows), two different bi-refraction cases with either positive-positive or positive-negative directions can be achieved. It is also worth mentioning that, in selected frequency ranges, the EFC is square-like therefore suggesting that the PC can produce self-collimation of an incoming diffused wave \cite{Ao}.

\begin{figure}[h!]
  \center{\includegraphics[scale=0.5]{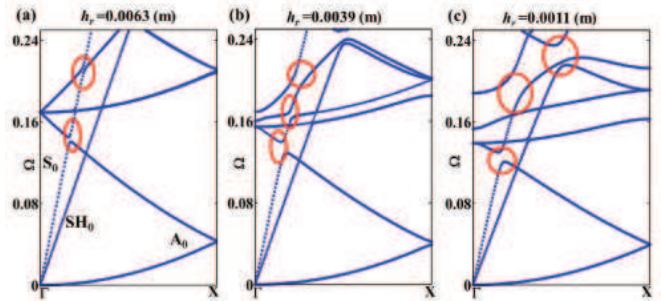}}
  \caption{Dispersion relations for different residual thickness values: (a) $h_r=$0.0063, (b) $h_r=$0.0039, and (c) $h_r=$0.0011.} \label{Fig5}
\end{figure}

The dispersion relations also reveal remarkable coupling between the different mode types highlighting the existence of a phenomenon known as \textit{mode hybridization} \cite{Pichard}. By inspecting the modal displacement of the fundamental non-monotonous mode along the $\Gamma-X$ direction, we observe that the branch to the left of the ZGVP (red solid line) is a dilatational $S_0$ mode while the branch on the right hand side is a flexural $A_0$ mode (green solid line). The change in the mode structure occurs very rapidly in the neighborhood of the zero group velocity point (ZGVP).
It is interesting to study how these modes develop and how they are affected by the ABH parameters, namely the residual thickness and taper exponent. Figure \ref{Fig5} shows the evolution of the dispersion relations with the residual thickness. Figure \ref{Fig5}(a), (b), and (c) correspond to different residual thickness cases where the coefficient $\varepsilon$ are set to 1, 3, and 5 meaning that the hole geometry vary from shallow to deep. Figure \ref{Fig5}(a) represents the shallow hole case ($h_r$ =0.00626) where the effects of the ABH should be less evident. As expected, the dispersion curves are quite similar to those of guided waves in a flat plate. In the low frequency range, we observe the $S_0$, $SH_0$ and $A_0$ modes and several folded branches of the $A_0$ mode due to the folding effect induced by the periodicity. The only significant difference is observed in correspondence to the mode crossing points. At these points, we observe splitting of the original modes and the generation of the non-monotonous hybrid modes connected by the ZGVP. The remaining part of the dispersion relations is essentially unaffected. By increasing the ABH slope (Figure \ref{Fig5}(b) and \ref{Fig5}(c)) the splitting and hybridization mechanisms become more evident even at higher frequency. A physical interpretation of this anomalous dispersion can be made in term of the coupled-wave theory \cite{AULD}. When two modes are coupled by a distributed mechanism (in our case the periodic inhomogeneity created by the ABH taper), significant interaction only occurs at synchronism, that is, near points where their dispersion curves cross. The coupling causes a characteristic splitting of the dispersion curves at the crossing point while, elsewhere, the modes are essentially unaffected.

\begin{figure}[h!]
  \center{\includegraphics[scale=0.5]{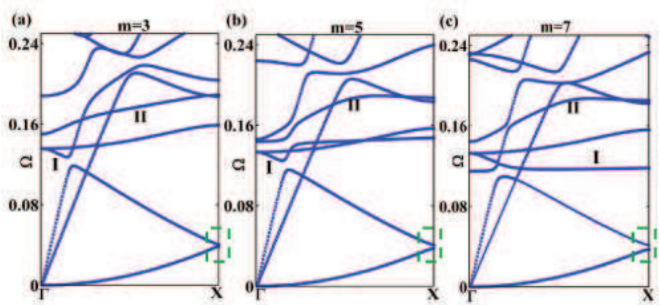}}
   \caption{Dispersion relation for different taper exponent values: (a) $m=$3, (b) $m=$5, and (c) $m=$7.} \label{Fig6}
\end{figure}

The other key design parameter is the exponential taper coefficient $m$. We consider a progression of the ABH profile from smooth (low $m$) to sharp (high $m$). For all cases the residual thickness is maintained constant at $h_r=0.0011$m by properly adjusting the coefficient $\epsilon$. The dispersion relations are shown in Figures \ref{Fig6}(a), (b), and (c) for $m=$3, 5, and 7. Of interest is the appearance of a bandgap between the fundamental flexural mode $A_0$ and the negative branch of the fundamental non-monotonous mode $A_{0f}$ (see green dashed box) following the splitting of the folded $A_0$ mode. This bandgap is considered to be related to the increased back-scattering occurring at large $m$ where the ABH smoothness criterion is no more satisfied \cite{Mironov,Feurtado}. Overall, the higher frequency modes (above the fundamental) are more evidently affected by the change in the taper exponent. As an example, mode I and II show substantial changes as $m$ increases. These two modes are characterized by monopole-like and dipole-like modal displacement fields inside the ABH and therefore are very sensitive to changes of the ABH profiles. Note also the formation (around $\Omega=0.1244$) of a nearly flat band (mode I). This flat mode suggests the existence of a deaf band which does not couple with any external wave \cite{C.T.Chan}.

In conclusion, we have introduced a class of 2D non-resonant single-phase phononic crystals made of a periodic lattice of Acoustic Black Holes. The remarkable dispersion characteristics of this material are even more surprising when considering the extremely simple design procedure that does not rely on the classical multi-phase material approach. Despite their outstanding simplicity, ABH PCs provide the same plethora of wave propagation effects typically observed in locally resonant materials, including negative refraction, bi-refraction and hybridization. They also present some peculiar phenomena, such as zero group velocity points on the fundamental Lamb modes and mode anisotropy, that were not observed before.

\bibliographystyle{apsrev}
\bibliography{Dispersion_ABH_ref}

\end{document}